\documentstyle[12pt]{article}
\begin{document}

\begin{center} {\bf DOES CHIRAL FERMION COUPLED TO A BACKGROUND
DILATON FIELD PRESERVE INFORMATION?}\\
Anisur Rahaman, Durgapur Govt. College, Durgapur - 713214, Burdwan, West Bengal, INDIA\\
e-mail: anisur.rahman@saha.ac.in \end{center}

\vspace{.5cm}

\vspace{2cm}
\begin{center} {\bf Abstract}
\end{center}

A model where chiral boson is coupled to a background dilaton
field is considered to study the s-wave scattering of fermion by a
back ground dilaton black hole. It is found that the scattering
process of chiral fermion does not violate unitarity and
information remains preserved. Faddevian anomaly plays a crucial
role on the information scenario.
\newpage

In recent years there has been a lot of interest in the physics of
information loss. Matter falling into the black holes carries some
information with it. That becomes inaccessible to the rest of the
world. A problem arises when the black hole evaporates through
Hawking radiation. It is a controversial issue whether or not
quantum coherence would be maintained during the formation and
subsequent evaporation of a black hole.
Quite a long ago Hawking suggested that the process did not
preserve information and unitarity failed to be maintained.
\cite{HAW}. It was an indication of a new level of
unpredictability in the realm of quantum mechanics induced by
gravity. There were plenty of opinions that went against Hawking's
suggestion. The main theme of those opinions was that the
information about the initial state of the system was carried by
some Plank scale steady remnant \cite{HOOFT, GARF}. However, it is
fair to admit that the issue gradually shifted against Hawking's
suggestion but it was not well settled.

In a recent publication, we find that Hawking has moved away from
his previous belief and suggested that quantum gravity interaction
does not lead to any loss of information. So there lies no problem
to maintain quantum coherence during the formation and subsequent
evaporation of the black hole \cite{HAW1}.
In spite of that, Hawking radiation effect on fermion information
loss problem is not well understood \cite{AHN}. Even now it has
been standing as a controversial issue.

This type of problem is very difficult to analyze in general.
However, some less complicated models are around us which were
solved to study this paradox \cite{GIDD, GIDD1, STRO, CALL, SUS}.
In these studies only the s-wave scattering of fermion incident on
the extremal charged black hole was considered. Angular momentum
coordinate becomes irrelevant in this situation and a two
dimensional effective action results. Though those simplified
models do not capture the detailed physics of black hole those
models contain the information loss paradox in a significant way
\cite {STRO, SUS, MIT}. Scattering of  fermion was not studied
only in connection with information loss, scattering of boson  was
also studied in \cite{FIS1, FIS2, FIS3} in the same context. The
scattering of chiral fermion off dilaton black hole was not
ignored too. It was considered with a particular interest in
\cite{MIT}.

The scattering of Dirac fermion itself is an interesting problem
\cite{GIDD, GIDD1, STRO, CALL, FIS1, FIS2, FIS3}. If Dirac
fermions are replaced by the chiral fermion it makes the analysis
more complex
because the chiral fermions generate anomaly in the energy
momentum sector when they couple to gravity \cite{BEL}. Therefore,
to get a solution for this type of complicated problem from the
study of the s-wave scattering of fermion from a back ground
dilaton field is interesting in its own right.

In ref. \cite{MIT}, the authors showed that the scattering of
chiral fermion can be studied in presence of the anomalies too if
one deals with the bosonized lagrangian. The process of
bosonization forces anomaly to enter into the picture. The anomaly
involved in studying this scattering problem \cite{JR} lead to the
author to conclude that information loss stood as a genuine
problem. Consequently, unitarity fails to be maintained. Of course
it was the correct conclusion, and the result was consistent too
with the Hawking's previous belief. Dirac fermion in \cite{STRO,
SUS} gave a completely opposite result. It exhibited an
information preserving result. In this situation one automatically
may be moved by the question 'Is there any possibility of getting
a result for chiral fermion which would be in agreement with the
Hawking's recent suggestion?'
The answer certainly may come from the investigation whether
chiral fermion can offer information preserving result like Dirac
fermion.
This motivates us to investigate the scattering of chiral fermion
off dilatonic black hole in a different setting.

To this end we consider a model where chiral fermion gets coupled
to a background dilaton field $\Phi$. Of course, electromagnetic
background is there. For sufficiently low energy incoming fermion,
the scattering of s-wave fermion incident on a charge dilaton
black hole can be described by the action
\begin{equation}
{\cal S}_f = \int d^2x[i\bar\psi\gamma^\mu[\partial_\mu +
ieA_\mu]\psi - {1\over 4} e^{-2\Phi(x)}F_{\mu\nu}F^{\mu\nu}].
\label{EQ1}
\end{equation}
Here e has one mass dimension. The indices $\mu$ and $\nu$ takes
the values $0$ and $1$ in $(1+1)$ dimensional space time. The
dilaton field $\Phi$ stands as a non dynamical back ground and its
only role in this model is to make the coupling constant a
position dependent one. Let us now define $g^2(x) = e^{2\Phi(x)}$.
Here as usual we will choose a particular dilaton background
motivated by the linear dilatonic vacuum of $(1+1)$ dimensional
gravity. Therefore, $\Phi(x) = -x^1$, where $x^1$ is space like
coordinate. The region $x^1 \to\+ \infty$, corresponds to exterior
space where the coupling $g^2(x)$ vanishes and the fermion will be
able to propagate freely. However, the region where $x^1 \to
-\infty$, the coupling constant will diverge and it is analogous
to infinite throat in the interior of certain magnetically charged
black hole.

The equation (\ref{EQ1}) is obtained from the action
\begin{equation}
S_{AF} = \int d^2\sigma\sqrt{g}[R + 4(\nabla\phi)^2 + {1\over
{Q^2}} - {1\over 2}F^2 + i \bar\psi D\!\!\!/\psi] \label{EQ2}
\end{equation}
for sufficiently low energy incoming fermion and negligible
gravitational effect \cite{STRO}. It is a two dimensional
effective field theory of dilaton gravity coupled to fermion. Here
$\Phi$ represents the scalar dilaton field and $\psi$ is the
charged fermion. Equation (\ref{EQ2}) was derived viewing the
throat region of a four dimensional dilatonic black hole as a
compactification from four to two dimension \cite{GARF, GIDD,
STRO}. Note that, in the extremal limit, the geometry is
completely non-singular and there is no horizon but when a low
energy particle is thrown into the non-singular extremal black
hole, it produces a singularity and an event horizon. In this
context, we should mention that the geometry of the four
dimensional dilatonic black hole consists of three regions
\cite{GARF, GIDD, GIDD1, STRO}. First one is the asymptotically
flat region far from the black hole. As long as one proceed nearer
to the black hole the curvature begins to rise and finally enters
into the mouth region (the entry region to the throat). Well into
the throat region, the metric is approximated by the flat two
dimensional Minkosky space times the round metric on the two
sphere with radius Q and equation (\ref{EQ2}) results. The dilaton
field $\Phi$ indeed increases linearly with the proper distance
into the throat.

In the present situation we are interested in studying the
scattering of chiral fermion. So we need to replace the vector
interaction by the chiral interaction which leads to
\begin{equation}
{\cal S}_f = \int d^2x[i\bar\psi\gamma^\mu[\partial_\mu +
ieA_\mu(1+\gamma_5)]\psi - {1\over 4}
e^{-2\Phi(x)}F_{\mu\nu}F^{\mu\nu}]. \label{EQQ}
\end{equation}
Equation (\ref{EQQ}), is the quantum chiral electrodynamics in
place of quantum electrodynamics with a dilaton back ground field
$\Phi$. This can be decoupled into the following
\begin{eqnarray}
{\cal S}_f &=& \int d^2x[\bar\psi_R\gamma_\mu\partial^\mu\psi_R+
i\bar\psi_L\gamma^\mu(\partial_\mu + ieA_\mu)\psi_L \nonumber \\
&-& {1\over 4} e^{-2\Phi(x)}F_{\mu\nu}F^{\mu\nu}].
\end{eqnarray}
Here $\psi_R$ represents the right handed fermion. This right
handed fermions remain uncoupled in this type of chiral
interaction. Integration over the right handed part gives a field
independent counter part which can be absorbed within the
normalization and the action reduces to the following
\begin{equation}
{\cal S}_f = \int d^2x[
i\bar\psi_L\gamma^\mu(\partial_\mu + i2e\sqrt{\pi}A_\mu)\psi_L \nonumber \\
- {1\over 4} e^{-2\Phi(x)}F_{\mu\nu}F^{\mu\nu}].\label{EQ5}
\end{equation}
In equation (\ref{EQ5}) e is replaced by $2e\sqrt{\pi}$ for later
convenience.
We now  bosonize the theory. The advantage of using the bosonized
version is that the anomaly automatically gets incorporated within
it. So the tree level bosonized theory contains the effect of
anomaly too. In order to bosonize the theory when we integrate out
the left handed fermion anomaly enter into the theory.  The
anomaly considered in this situation is of Faddeevian class
\cite{FADDEEV}.
 With the generalized
Faddeevian anomaly \cite{MG, AR1} the bosonized action reads
\begin{eqnarray}
{\cal L}_{CH} &=& {1\over 2}\partial_\mu\phi\partial^\mu\phi +
e(\eta^{\mu\nu} - \epsilon^{\mu\nu})
\partial_\mu\phi A_\nu \nonumber \\
&+& {1\over 2} e^2[A_0^2 - A_1^2 +2 \alpha A_1( A_0 + A_1)] -
{1\over 4} e^{-2\Phi(x)}F_{\mu\nu}F^{\mu\nu} .\end{eqnarray} This
equation though shows no Lorentz covariant structure it has the
physical Lorentz invariance \cite{AR1}. Here we impose the chiral
constraint $\Omega(x) = \pi(x) - \phi'(x)$, to express the action
in terms of chiral boson following the procedure available in
\cite{KH}. In terms of chiral boson the model turns into
\begin{eqnarray}
{\cal L}_{CH} &=& \dot\phi\phi' -\phi'^2 + 2e(A_0 - A_1)\phi'\nonumber \\
&+& {1\over 2}e^2 [2(\alpha  - 1)A_1^2 + 2(\alpha + 1)A_0A_1] +
{1\over 2} e^{-2\Phi(x)}F_{01}^2. \label {LCH}
\end{eqnarray}
Here $\phi$ represents a scalar field. Note that $\dot\phi^2$ is
absent because the first two terms in the lagrangian (\ref{LCH})
corresponds to the kinetic term for chiral boson \cite{SIG, FLO}.
It is now necessary to carry out the Hamiltonian analysis of the
theory to observe the role of dilaton field on the equation of
motion. From the standard definition of momentum the canonical
momenta corresponding to the chiral boson field $\phi$, the gauge
field $A_0$ and $A_1$ are obtained.
\begin{equation}
\pi_\phi = \phi',\label{MO1}
\end{equation}
\begin{equation}
\pi_0 = 0,\label{MO2}
\end{equation}
\begin{equation}
\pi_1 = e^{-2\phi(x)}(\dot A_1 - A_0')={1\over {g^2}}(\dot A_1 -
A_0).\label{MO3}
\end{equation}
Here $\pi_\phi$, $\pi_0$ and $\pi_1$ are the momenta corresponding
to the field $\phi$, $A_0$ and $A_1$. Using the above equations it
is straightforward to obtain the canonical Hamiltonian through a
Legendre transformation. The canonical Hamiltonian is found out to
be
\begin{eqnarray}
H_C &=& \int dx[{1\over 2} e^{2\Phi}\pi_1^2 + \pi_1A_0' + \phi'^2
-
 2e(A_0 - A_1)\phi' \nonumber \\
&-& {1\over 2}e^2[2(\alpha -1)A_1^2 + 2(1 +
\alpha)A_0A_1)]].\label{CHAM}
\end{eqnarray}
The Hamiltonian though acquires an explicit space dependence
through the dilaton field $\Phi(x)$, it has no time dependence. So
it is preserved in time. Equation (\ref{MO1}) and (\ref{MO2}) are
the primary constraints of the theory. Therefore, it is necessary
to write down an  effective Hamiltonian:
\begin{equation}
H_{eff} = H_C + u\pi_0 + v(\pi_\phi - \phi'),
\end{equation}
where $u$ and $v$ are two arbitrary Lagrange multipliers. The
primary constraints (\ref{MO1}) and (\ref{MO2}) have to be
preserve in order to have a consistent theory. The preservation of
the constraint (\ref{MO2}), leads to a new constraint which is the
Gauss law of the theory:
\begin{equation}
G = \pi_1' + 2e\phi' +  e^2(1 + \alpha)A_1 = 0. \label{GAUS}
\end{equation}
The preservation of the constraint (\ref{MO1}) though does not
give rise to any new constraint it fixes the velocity $v$ which
comes out to be
\begin{equation}
v = \phi' - e(A_0 - A_1). \label{VEL}
\end{equation}
The constraint (\ref{GAUS}), also has to be conserved and the
conservation of it requires
\begin{equation} \dot G = 0\label{CON}
.\end{equation} A new constraint
\begin{equation}
(1 + \alpha)e^{2\Phi}\pi_1 + 2\alpha (A_0' + A_1') =
0,\label{FINC}
\end{equation}
appears from the preservation condition (\ref{CON}). No new
constraints comes out from the preservation of (\ref{FINC}). So we
find that the phase space of the theory contains the following
four constraints.
\begin{equation}
\omega_1 = \pi_0, \label{CON1}
\end{equation}
\begin{equation}
\omega_2 = \pi_1' + 2e\phi' +  e^2(1 + \alpha)A_1 = 0,\label{CON2}
\end{equation}
\begin{equation}
\omega_3 = (1 + \alpha)e^{2\Phi}\pi_1 + 2\alpha (A_0' + A_1') =
0,\label{CON3}
\end{equation}
\begin{equation}
\omega_4 = \pi_\phi - \phi'. \label{CON4}
\end{equation}
The four constraints (\ref{CON1}), (\ref{CON2}), (\ref{CON3}) and
(\ref{CON4}) form a second class set and all of these are weak
condition up to this stage. If we impose these constraints
strongly into the canonical Hamiltonian (\ref{CHAM}), the
canonical Hamiltonian gets simplified into the following.
\begin{equation}
H_R = \int dx^1[ {1\over 2}e^{2\Phi(x)}\pi_1^2 + {1\over {4e^2}}
\pi_1'^2 + {1\over 2}(\alpha - 1) \pi_1'A_1 + {1\over 4}e^2[(1 +
\alpha)^2 - 8\alpha)]A_1^2]. \label{RHAM}
\end{equation}
$H_R$ given in equation (\ref{RHAM}), is generally known as
reduced Hamiltonian. According to Dirac, Poisson bracket gets
invalidate for this reduced Hamiltonian \cite{DIR}. This reduced
Hamiltonian however remains consistent with the Dirac brackets
which is defined by
\begin{equation}
[A(x), B(y)]^* = [A(x), B(y)] - \int[A(x), \omega_i(\eta)]
C^{-1}_{ij}(\eta, z)[\omega_j(z), B(y)]d\eta dz, \label{DEFD}
\end{equation}
where $C^{-1}_{ij}(x,y)$ is defined by
\begin{equation}
\int C^{-1}_{ij}(x,z) [\omega_j(z), \omega_k(y)]dz =\delta(x-y)
\delta_{ik}. \label{INV}
\end{equation}
Here $i$ and $j$ runs from $1$ to $4$ and $\omega$'s represent the
constraints of the theory. With the definition (\ref{DEFD}), we
can compute the Dirac brackets between the fields describing the
reduced Hamiltonian $H_R$. The Dirac brackets between the fields
$A_1$ and $\pi_1$ are required to obtain the theoretical spectra
(equation of motion):
\begin{equation}
[A_1(x), A_1(y)]^* = {1\over {2\alpha e^2}}\delta'(x-y),
\label{DR1}
\end{equation}
\begin{equation}
[A_1(x), \pi_1(y)]^* = {(\alpha -1)\over
{2\alpha}}\delta(x-y),\label{DR2}
\end{equation}
\begin{equation}
[\pi_1(x), \pi_1(y)]^* = -{(1+\alpha)^2 \over
{4\alpha}}e^2\epsilon(x-y).\label{DR3}
\end{equation}
Using (\ref{RHAM}), (\ref{DR1}), (\ref{DR2}) and (\ref{DR3}), we
obtain the following first order equations of motion:
\begin{equation}
\dot A_1 = {(\alpha-1) \over {2\alpha}} e^{2\Phi(x)}\pi_1 - A_1',
\label{EQM1}
\end{equation}
\begin{equation}
\dot \pi_1 = \pi_1' + 2(\alpha - 1)e^2A_1. \label{EQM2}
\end{equation}
After a little algebra we find that the field $\pi_1$ satisfy a
Klein Gordon equation
\begin{equation}
(\Box - e^{2\Phi}e^2{(\alpha-1)^2\over {\alpha}})\pi_1 = 0
\label{SPEC},
\end{equation}
The equation (\ref{SPEC}), represents a massive boson with square
of the mass $m^2 = g^2{{-(1-\alpha)^2}\over \alpha }e^2$. Here
$\alpha$ must be negative in order to have the mass of the boson a
physical one. Mass of this boson however in this particular
situation is not constant. It contains a position dependent factor
$g^2=e^{2\Phi}$ where $\Phi = -x^1$ for the background motivated
by the linear dilatonic vacuum of $(1+1)$ dimensional gravity.
Therefore, $m^2 \to \ + \infty$ when $x^1 \to\ - \infty$ and $m^2
\to \ 0$ when $x^1 \to\+ \infty$. Thus mass of the boson is found
to be  increased indefinitely in the negative $x^1$ direction
which implies that any finite energy contribution must be totally
reflected and an observer at $x^1 \to \infty$ will recover all
information. To be more precise, mass will vanish near the mouth
(the entry region to the throat) but increases indefinitely as one
goes into the throat because of the variation of this space
dependent factor $g^2$. Since massless scalar is equivalent to
massless fermion in $(1+1)$ dimension, we can conclude that a
massless fermion proceeding into the black hole will not be able
to travel an arbitrarily long distance and will be reflected back
with a unit probability. So, there will be no information loss and
a unitary s-matrix can be constructed for this particular
scattering problem. This results reminds us the scattering of
Dirac fermion \cite{STRO, SUS}. Thus in the description of
scattering of chiral fermion where $U(1)$ anomaly has been taken
into account with the introduction of Faddeevian class of
anomalous term is found to be free from the {\it dangerous}
information loss problem.  Note that this result is just opposite
to the conclusion of \cite{MIT}.
The analysis available in \cite{MIT} and the present analysis
differs only in the anomaly structure but this particular class of
anomaly has brought a novel change in  the information scenario
for chiral fermion. It is tempting to note that the change
appeared in the scattering of chiral fermion in connection with
information  scenario is consistent with the standard belief as
well as with the Hawking's recent suggestion. It is fair to admit
here that the mechanism how this new setting of anomaly saved the
model from the danger of information loss is not clear and
certainly needs further investigation.
However, one should not think that it has come out of the blue. We
observed the crucial role of anomaly elsewhere too, e. g., in
quantum electro dynamics and quantum chiral electrodynamics etc.
\cite{JR, KH, PM, MG, AR1}. A famous instance is the removal of
the long suffering of the chiral electro dynamics from the non
unitarity problem \cite{JR}.
To conclude this we would like to mention that anomaly not always
appears as a disturbing term, some times it appears as a surprise
to give a relief from disturbance. The present work  would be an
an example of that.

\noindent {\bf Acknowledgment}: It is a pleasure to thank the
Director, Saha Institute of Nuclear Physics and the Head of the
Theory Group of Saha Institute of Nuclear Physics, Kolkata for
providing working facilities. I would also like to thanks the
referee for his valuable suggestion.

\end{document}